\documentclass[twocolumn,amsmath,amssymb,prl]{revtex4}
\usepackage{graphicx}
\usepackage{dcolumn}
\usepackage{bm}

\begin{document}
\input{epsf}

\def\apj{Astrophys. J.}
\def\mnras{Mon. Not. R. Astr. Soc.}
\def\aap{Astr. \& Astrophys.}
\def\apjl{Astrophys. J. Lett.}
\def\apjs{Astrophys. J. Supp.}
\def\physrep{Phys. Rep.}

\newcommand{\la}{\lesssim}
\newcommand{\ga}{\gtrsim}

\title[Evaporation of X-ray Clusters]{Thermal Evaporation of Gas from X-ray
Clusters} 

\author{Abraham Loeb} 
\affiliation{Astronomy Department, Harvard University, 60 Garden Street,
Cambridge, MA 02138}

\begin{abstract}

A fraction of the thermal protons in the outer envelope of an X-ray cluster
have velocities that exceed the local escape speed from the cluster
gravitational potential. The Coulomb mean-free-path of these protons is
larger than the virial radius of the cluster at temperatures $\gtrsim
2.5$keV.  The resulting leakage of suprathermal particles generates a
collisionless shock in neighboring voids and fills them with heat and
magnetic fields. The momentum flux of suprathermal particles cannot be
confined by magnetic tension at the typical field strength in the periphery
of cluster halos ($\ll \mu$G).  Over a Hubble time, thermal evaporation
could drain up to a tenth of the cluster gas at its virial temperature.
The evaporated fraction could increase dramatically if additional heat is
deposited into the gas by cluster mergers, active galactic nuclei or
supernovae.  Thermal evaporation is not included in existing cosmological
simulations since they are based on the fluid approximation. Measurements
of the baryon mass fraction in the outer envelopes of hot clusters
(through their Sunyaev-Zel'dovich effect or X-ray emission) can be used to
empirically constrain their evaporation rate.

\end{abstract}


\maketitle

\section{I. Introduction}

It is commonly assumed that the fluid approximation applies to the
virialized gas in X-ray clusters. However, the Coulomb mean-free-path of a
proton or an electron in a plasma of temperature $T$ and density contrast
$\Delta$ relative to the present-day ($z=0$) mean cosmic density is given
by,
\begin{equation}
\lambda (0)= 1.3 {\rm Mpc} \left({k_BT \over 5{\rm keV}}\right)^{2} 
\left({\Delta\over 25}\right)^{-1} . 
\label{eq:1}
\end{equation}
A value of $\Delta\sim 25$ characterizes the local gas density at the
virial radius $r_{\rm vir}$ of cluster dark--matter halos
\citep{NFW,Fab}. For a cluster that virializes at the present time
\citep{Loeb},
\begin{equation}
r_{\rm vir} (0)=2.3 {\rm Mpc}\left({k_BT\over 5~{\rm keV}}\right)^{1/2},
\label{eq:2}
\end{equation}
where we adopt the standard set of cosmological parameters \citep{WMAP} and
assume primordial composition with 24\% helium by mass.  Thus, the
mean-free-path of protons and electrons is comparable to the size of the
virialized envelope of hot X-ray clusters. Hydrodynamic simulations
\citep{Norman,SFR,Ettori,Evrard} show that a substantial portion of the
surface of these envelopes is in contact with rarefied voids in which
$\Delta<1$. Protons moving away from the cluster envelope towards a
neighboring void must be confined by gravity since Coulomb collisions in
the surrounding $\Delta<1$ environment are too rare to confine them. This
applies even better to protons with a kinetic energy $E$ larger than the
mean thermal value, since the Coulomb mean-free-path scales as $E^{2}$. If
the kinetic energy of a proton exceeds the local gravitational binding
energy, gravity cannot keep the proton bound to the cluster.  Thermal
plasma would therefore evaporate from the cluster (as each escaping proton
must be accompanied by an electron so as to preserve charge
quasi-neutrality). This process reflects a departure from the fluid
description of the intracluster gas and cannot be captured by numerical
hydrodynamic simulations which are based on the fluid approximation. The
purpose of the this paper is to gauge the significance of this evaporation
process in X-ray clusters and to motivate future numerical simulations
involving kinetic theory in cosmological three-dimensional
geometries. Significant evaporation would modify the baryon fraction in
clusters and weaken the case for it having a universal value that can be
used for cosmological distance determination
\citep{Ettori,Pen,Allen,Carl,Hall}.  The evaporation would also heat the
gas in voids \cite{Loe} and potentially change the Ly$\alpha$ absorption
signature of voids in quasar spectra at low redshifts \cite{Dave}.  The
fractional helium abundance in clusters should also increase as only
hydrogen evaporates.

The use of X-ray clusters for cosmological parameter determination
(e.g. regarding the dark energy equation of state) is one of the 
active frontiers in cosmology. A large number of surveys for the
Sunyaev-Zeldovich effect as well as the X-ray emission of clusters are
currently being conducted. The observational data is usually compared to
cosmological hydrodynamical simulations. In this paper we point out that
the fluid approximation is not valid in the outer parts of hot clusters,
where thermal evaporation of gas is likely to take place. The evaporation
needs to be included in future studies attempting to refine the use of
X-ray clusters for precision cosmology.

\section{II. Mass Loss Rate} 

The escape velocity $v_{\rm esc}$ from a virialized cluster is defined as
the velocity necessary to overcome the gravitational potential energy per
unit mass $W$ of the cluster, ${1\over 2} v_{\rm esc}^2=-W({\bf r})$.  The
virial theorem relates the mass-averaged value of $W$ to twice the mean
kinetic energy per unit mass in the cluster $\langle -W\rangle=\langle
v^2\rangle$ and implies $\langle v_{\rm esc}^2\rangle =4 \langle
v^2\rangle$, where angular brackets denote mass-averaged values
\citep{Binney-Tremaine}.

In the envelopes of clusters, the timescale for equilibration of the
electron and proton temperatures through Coulomb collisions is comparable
to the Hubble time \citep{Fox-Loeb,Tak}.  However, within any plasma
confined by a gravitational field, there is an electric field which
balances the pressure gradient of the electrons \citep{Loeb88},
\begin{equation}
{\bf { {E}}} = -{{\bf{\nabla}} p_e\over e n_e},
\end{equation}
where $p_e=n_e k_BT_e$ is the electron pressure and $e$ is the electron
charge.  This electric field binds the electrons, which are otherwise too
light to be bound by gravity. The same field reduces the depth of the
gravitational barrier for a proton to escape. In an electron-proton plasma
with a single temperature, the gravitational potential barrier is reduced
by a factor of 2.  If $T_e=0$ then $E=0$ and the barrier equals the full
gravitational potential. The evaporation rate depends only on the mean
thermal energy per electron-proton pair irrespective of how they share this
energy. For simplicity, we will assume that all particles have the same
temperature from now on.

The local barrier in the outer envelope of the cluster is smaller than its
mass-averaged value across the entire cluster.  The evaporation rate
depends on the local value of the parameter, $\eta ({\bf {r}})\equiv v_{\rm
esc}^2/ {\bar{v^2}}$, at the envelope.  For a Maxwellian velocity
distribution of an equilibrium temperature $T$, the average kinetic energy
per proton is ${1\over 2} m_p {\bar{v^2}} = {3\over
2}k_BT$, where $m_p$ is the proton mass.  The fraction $\epsilon({\bf r})$
of the thermal particles that exceed the escape speed is given by
\begin{equation}
\epsilon={2\over \sqrt{\pi}} \int_{3\eta/2}^{\infty}\exp({-\zeta})
\zeta^{1/2} d\zeta .
\end{equation}
For $1\lesssim \eta\lesssim 4$, we find $\epsilon\approx 1.47 e^{-1.3\eta}$
to better than $10\%$.  As particles from the tail of the Maxwellian
distribution escape the cluster, this tail is re-populated through Coulomb
collisions in the higher density gas bordering the envelope from within
the cluster.

Next we calculate the fraction of protons above the escape speed as a
function of fractional radius $R=(r/r_{\rm vir})$.  Hydrostatic equilibrium
for a primordial composition implies
\begin{equation}
{k_BT(r)\over m_p}={1.18\over 2\rho_{\rm g}(r)}\int_r^{\infty}
{GM(r')\rho_{\rm g}(r')\over r'^2} dr' ,
\label{eq:T}
\end{equation}
and the escape speed for a proton (after subtracting off the repulsive
electric force from the gravitational force) is
\begin{equation}
v_{\rm esc}^2(r)=0.82\int_r^\infty {GM(r')\over r'^2}dr',
\end{equation}
where $\rho_{\rm g}(r)$ is the gas mass density and $M(r)$ is the total
interior mass at radius $r$.  Numerical simulations and X-ray data indicate
that the hot gas follows the dark matter distribution in the outer parts of
X-ray clusters \citep{Komatsu}. We therefore adopt the NFW \citep{NFW}
density profile, for which $M(x)\propto [\ln(1+x)-x/(1+x)]$ and $\rho_{\rm
g}\propto 1/[x(1+x^2]$, where $x=r/r_s$ and $r_s=r_{\rm vir}/C$ with $C$
being the concentration parameter.  The normalization of the virial mass or
virial radius scale out from the expression for $\eta=v_{\rm
esc}^2/[3k_BT/m_p]$, which depends only on $R$ and $C$.  The radial profile
of the gas temperature is calculated from Eq. (\ref{eq:T}).

Figure 1 depicts $\epsilon(R)$ for three values of $C$. We use the
integration upper limit of $r'=10r_{\rm vir}$ instead of $\infty$ in the
above integrals, as it represents the typical radius of a void or half the
distance between rich clusters.  For the characteristic value of recently
collapsed halos, $C\sim 4$, the escape fraction is $\sim 7.6\%$ at $R\sim
1$, corresponding to $\eta \sim 2.3$
\footnote{Note that the gas responds to the gravitational potential of the
dark matter and need not possess $\langle \eta\rangle=4$ as expected from
the virial theorem for self-gravitating matter.  In particular, the factor
of ${1\over 2}$ that is usually inserted to avoid double counting in the
total gravitational interaction energy, need not be included for test
particles in an external potential.  }.  The asymptotic power-law slope of
$-3$ for the radial density profile yields $\eta =1.87$ and $\epsilon
\approx 13\%$ for large values of $C$.  Since $\rho_{\rm g}\propto R^{-3}$
at $R\gg 1/C$, fresh suprathermal particles can be supplied to the
collisionless envelope from the highly collisional interior at $R\lesssim
0.5$.

Clearly, the cluster plasma has to be sufficiently hot in order for the
suprathermal particles not to be reflected back into the cluster through
Coulomb collisions. Their mean-free-path ($\propto \eta^2$) exceeds the
virial radius of a cluster for temperatures
\begin{equation}
k_BT> 2.5~{\rm keV}\times (1+z) \left({\eta\over
2.3}\right)^{-4/3}\left({\Delta \over 25}\right)^{2/3} ,
\label{eq:cond}
\end{equation}
where we used Eqs. (\ref{eq:1}) and (\ref{eq:2}) with the additional
scaling to arbitrary redshift, $z$.

\begin{figure} [ht]
\centerline{\epsfxsize=3.4in \epsfbox{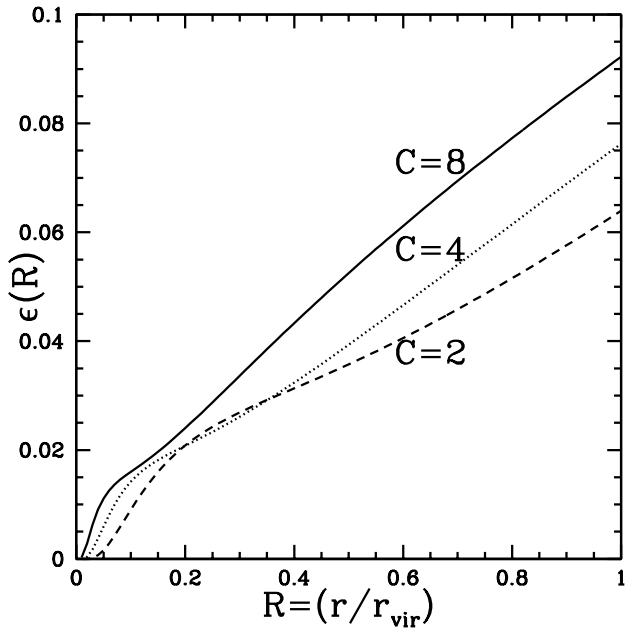}}
\caption{Fraction $\epsilon$ of protons above the escape speed as a
function of fractional radius $R=(r/r_{\rm vir})$ for an NFW \citep{NFW}
density profile with different values of the concentration parameter $C=2$
(dashed line), $4$ (dotted) and $8$ (solid). }
\label{figure1}
\end{figure}

Hydrodynamic simulations \citep{Norman,SFR,Ettori,Evrard} indicate that
infalling gas is often channeled into clusters through filaments, while
other parts of the cluster border low-density voids.  Thermal evaporation
will occur preferentially across the contact surface with voids where there
is negligible resistance to the outward flux of escaping
particles. Magnetic fields which are tangled on small scales would suppress
evaporation since the proton Larmor radius is negligibly small. Such fields
could originate either from galactic winds \citep{Ens} and quasar outflows
\citep{Fur,Kron} or be generated in the collisionless shocks that virialize
the cluster gas \citep{Keshet,Kuls}. The fields are likely to be
accompanied by relativistic cosmic-rays which are produced by the same
sources. Indeed merging X-ray clusters are known to be surrounded by radio
halos due to synchrotron emission by relativistic electrons
\citep{Govoni,Keshet}. The cosmic-ray pressure might subject the magnetic
field to the Parker instability \citep{Shu}, producing channels of open
field lines along which thermal evaporation will not be inhibited.  In
deriving the evaporation rate below, we include a suppression factor $f<1$
to account for the fraction of the cluster surface area where evaporation
is not suppressed by inflowing gas or by magnetic fields.

The mass evaporation rate of cluster plasmas that satisfy
Eq. (\ref{eq:cond}) can be estimated from the {\it outward} flux of all
thermal particles above the escape speed of the cluster at the virial
radius \footnote{In the general case of a relaxed gas distribution with an
arbitrary temperature profile, the evaporation rate is given by the flux of
escaping particles at the radius $r_{\rm ev}$ where $\lambda (r_{\rm
ev})\sim r_{\rm ev}$, and the evaporation will be suppressed if no such
radius exists.},
\begin{equation}
{\dot M_{\rm g}} = 4\pi\times 0.76 f \epsilon \rho_{\rm g}(r_{\rm vir})
r_{\rm vir}^2 {v_{\rm esc}} \approx 3 f\epsilon\eta^{1/2} \left( {\Delta
\over 25}\right) \left({M_{\rm g}\over t}\right) ,
\label{eq:mdot}
\end{equation}
where $M_{\rm g}$ is the interior gas mass at the virial radius and
$t=H(z)^{-1}$ is the Hubble time at the redshift of interest.  Hence, the
fraction of the cluster gas that evaporates over a fraction $\alpha$ of the
Hubble time is $\sim 3\alpha f\epsilon \eta^{1/2}$. For $\alpha f\sim
{1/3}$ and $\eta\sim 2.3$, the total evaporated fraction is $\sim 10\%$,
comparable to the fraction of baryons which are locked in stars
\citep{Hogan}. A deficit larger by a factor of a few (attainable with
$f\sim 1$ and $\eta\sim 2$) is still allowed by existing data on the baryon
mass fraction in cluster envelopes \citep{Vikhlinin,Allen,Fab,Carl}.  As
accretion will taper off in the future of a $\Lambda$--dominated universe
\citep{Nagamine,Busha}, the hot envelopes of X-ray clusters might be
depleted significantly over a timescale $\gtrsim 10^{11}$yr.

In order for magnetic confinement to operate, the magnetic tension
$B^2/8\pi$ needs to exceed the momentum flux of the suprathermal particles
$\sim \epsilon\rho_{\rm g} v_{\rm esc}^2$, 
or else the magnetic field will be moved around by the suprathermal wind.
Although the field is anchored to the bulk of the plasma, a large load of
suprathermal particles will reshape its topology and allow escape,
e.g. through the ``garden-hose'' instability \citep{Tidman,Hall} and other
buckling instabilities of the magnetic field \citep{Parker} that will open
channels of outflow for the suprathermal particles.  These instabilities
are sourced by the anisotropic distribution of particles in the local
plasma frame and are particularly powerful when the pressure anisotropy of
the plasma exceeds the magnetic pressure.  The required field strength to
resist the outward momentum flux of suprathermal particles is
\begin{equation}
B\gtrsim 1 \mu{\rm G}~ \left[\left({\epsilon\over 0.1}\right)
\left({\eta \over 2.3}\right) \left({\Delta \over 25}\right)\left({k_BT\over
5~{\rm keV}}\right)\right]^{1/2} .
\label{eq:mom}
\end{equation}
Synchrotron and hard X-ray observations as well as Faraday rotation
measurements in the {\it periphery} of the extended halos around clusters
infer a magnetic field strength which is 3--10 times lower than the
required magnitude \citep{Rephaeli,Xu,Kronberg}.  The associated magnetic
tension is at least an order of magnitude smaller than required and could
therefore allow significant leakage of suprathermal particles.

The mean-free-path of suprathermal protons might be reduced through
scattering on plasma waves \citep{Stix,Howes}, which are excited in the
intracluster plasma by the collisionless shock formed when accreting gas
impacts on the cluster boundary or through the turbulence generated by the
motion of sub-halos within the cluster.  However, the excitation of plasma
turbulence is expected to be unsteady and the waves would leak out during
quiescent episodes of weak accretion.  The long mean-free-path of protons
makes a large viscosity coefficient that would have suppressed acoustic
turbulence altogether in the absence of magnetic fields.  (Although
existing hydrodynamic simulations identify acoustic turbulence in cluster
cores, their limited numerical resolution and incomplete treatment of
plasma physics do not allow them to examine reliably the turbulence damping
rate \citep{Dol,Vazza}.)  Plausible excitation mechanisms of Alfven waves
in a relaxed cluster produce negligible power on wavelengths that resonate
with the microscopic ion gyro-radius, $\sim 10^{11}~{\rm cm}(B/0.1\mu{\rm
G})^{-1}$; under such circumstances the waves cannot scatter effectively
the suprathermal particles and suppress their large momentum flux.

The outgoing beam of suprathermal particles would generate a shock in
surrounding voids through the plasma two-stream instability \citep{Tidman}.
The dispersion relation of the instability at finite temperatures implies
that a Maxwellian-tail beam will not excite this instability within the hot
cluster plasma but only as it enters the much colder environment of a void.
The resulting shock will heat the gas in the voids and enrich
it with magnetic fields.

\section{III. Discussion}

We have shown that if magnetic confinement is ineffective
(Eq. \ref{eq:mom}), then thermal evaporation is capable of removing up to a
tenth of the cluster gas at its virial temperature over a Hubble time
(Eq. \ref{eq:mdot}).  Local energy injection from cluster mergers
\citep{Gomez}, active galactic nuclei \citep{Pet} or supernovae
\citep{Bialek,Ettori} could enhance the evaporated fraction as it would
raise the gas temperature and lower the corresponding value of $\eta$. The
evaporated fraction is exponentially sensitive to enhancements in the gas
temperature, with $\epsilon \propto \exp(-1.3\eta)$. For a fixed energy
injection per baryon (as expected from a constant star/black-hole formation
efficiency), the change in $\eta$ would be largest at low values of
$T$. Indeed, the baryon fraction in the shallower potential wells of poor
clusters appears to be smaller than in rich clusters \citep{Vikhlinin},
indicating gas loss through additional heat supply to the virial energy
budget.

An essential ingredient for the evaporation to be maintained is that the
tail of suprathermal protons will be replenished despite the low
collisionality of the plasma in the cluster atmosphere.  Indeed protons
just below the escape speed are turned around by gravity and then fall back
towards the cluster center. Along their trajectory they encounter higher
density gas, where they collide with other protons and re-populate the
Maxwellian tail once again. An approach to the collisional regime merely
requires sinking by a factor of 2 in radius for the NFW density profile,
where $\rho_g\propto r^{-3}$ for $r_{\rm vir}/C\lesssim r\lesssim r_{\rm
vir}$.  This replenishment process does not occur for weakly interacting
dark matter particles, which would therefore not evaporate steadily in
relaxed clusters. The dark matter might nevertheless experience transient
evaporation episodes as a result of time-dependent gravitational
interactions during violent mergers.

Although existing simulations of clusters are performed in the fluid
approximation, they can nevertheless be implemented for calculating the
thermal evaporation rate under more realistic conditions than those
considered in Fig. 1. The simulated distributions of the gravitational
potential and the gas temperature can be used to derive $v_{\rm esc}({\bf
r})$ and $\epsilon ({\bf r})$ within the region where the plasma is
collisionless.  A corresponding evaporation term can then be added to the
local conservation laws.  The significance of magnetic confinement could be
examined through magneto-hydrodynamic simulations \citep{Dolag,Br}, with
the important inclusion of the stress tensor for suprathermal particles and
cosmic-rays.

Better observational constraints on the intracluster magnetic field
\citep{Rephaeli,Xu,Kronberg,Gaens} could determine whether the inferred
magnetic tension in cluster halos is indeed unable to sustain the momentum
flux of suprathermal protons above the escape speed (Eq. \ref{eq:mom}).
The same fields affect intracluster cosmic-rays which produce neutral pions
that decay into $\gamma$-rays \citep{Miniati}.  Future $\gamma$-ray
observations with GLAST\footnote{http://glast.stanford.edu/} could
constrain the magnetic confinement time of cosmic-rays, and by
extrapolation -- suprathermal particles.

Existing X-ray data on the baryon mass fraction \citep{Vikhlinin,Allen,Fab}
probe only the interior of clusters ($R\lesssim 0.3$), where $\lambda<r$.
The observed baryon fraction there is already known to be smaller than the
expected cosmic average. Future extensions of existing data sets to the
low-density envelopes of hot clusters through deep observations of their
X-ray emission or Sunyaev-Zel'dovich effect \citep{Afshordi}, could
determine whether there is a deficit in the baryon fraction relative to the
cosmic value beyond the level expected from the consumption of gas by star
formation \citep{SFR}.

\bigskip
\paragraph*{Acknowledgments}
I thank A. Broderick, W. Forman, R. Kulsrud, J. Raymond, G. Rybicki, A.
Vikhlinin, and E. Waxman for useful comments on the manuscript, and
L. Chuzhoy for suggesting to consider the effect of helium.  This work was
supported in part by NASA grants NAG 5-1329 and NNG05GH54G and by Harvard
University grants.

\end{document}